\documentclass{article}
\usepackage{spconf,amsmath,graphicx}

\usepackage{booktabs}
\usepackage{subfig}
\usepackage{url}
\usepackage{cite}
\usepackage{amssymb}
\usepackage{mathtools, nccmath}
\usepackage{multirow}  
\usepackage{xcolor}         

\title{Towards Improved Room Impulse Response Estimation for Speech Recognition}
\address{$^1$ University of Maryland, College Park, USA,
         $^2$ Reality Labs Research at Meta, Redmond, WA USA}
\begin{document}
%
\maketitle
\begin{abstract}

We propose a novel approach for blind room impulse response (RIR) estimation systems in the context of a downstream application scenario, far-field automatic speech recognition (ASR).  We first draw the connection between improved RIR estimation and improved ASR performance, as a means of evaluating neural RIR estimators.  We then propose a generative adversarial network (GAN) based architecture that encodes RIR features from reverberant speech and constructs an RIR from the encoded features, and uses a novel energy decay relief loss to optimize for capturing energy-based properties of the input reverberant speech.  We show that our model outperforms the state-of-the-art baselines on acoustic benchmarks (by 17\% on the energy decay relief and 22\% on an early-reflection energy metric), as well as in an ASR evaluation task (by 6.9\% in word error rate).

\end{abstract}
\begin{keywords}
room impulse response, blind estimation
\end{keywords}
%

\section{Introduction}
\label{sec:intro}

The reverberation effects in audio recordings depend on the acoustic environment (e.g., room geometry, room materials, etc.) in which they are recorded. For instance, speech in large auditoria sounds perceptually very different from speech in small meeting rooms. Such reverberation effects (i.e., early reflections, late reverberation, the amount of sound scattering and diffraction, etc.) are characterized by a transfer function known as the room impulse response (RIR). Measuring or simulating accurate RIRs for a given acoustic environment is vital for a variety of speech and audio applications including speech recognition~\cite{irgan,butreverbDB}, speech dereverberation~\cite{dereverb1,dereverb2}, speech separation~\cite{seperate1,seperate2} and augmented/virtual reality~\cite{VR}.


Accurate measurement of RIRs is prohibitively expensive and requires expert guidance~\cite{butreverbDB}. 
On the other hand, simulating accurate RIRs requires 3D mesh representations of the underlying scenes~\cite{mesh2ir} and complete knowledge of material properties~\cite{soundspaces2,gwa}. Therefore, it is not practical to precisely simulate RIRs for a real-world environment without their 3D representation. As an alternative, we propose an approach to directly estimate the RIR from single-channel speech recordings. RIR estimation via blind system identification is a well studied problem in classical signal processing~\cite{lin2006bayesian, crammer2006room,  es2}. Recently, deep learning based end-to-end RIR estimation methods have been employed with some success \cite{FiNS}.

In this work, we focus on an RIR estimation framework from the point of view of a specific use case, automatic speech recognition (ASR). 
It is well known that reverberant speech inputs affect the performance of ASR systems in home voice assistants, particularly in the presence of a domain mismatch between the training and test data in terms of reverberation effects \cite{out_data}. Such a mismatch is a direct consequence of the need for augmenting training data to be similar to the test data by estimating the reverberation effects in the testing environment to ensure learnability when large neural networks are involved. In principle, one can utilize the findings from recent RIR estimation models like the filtered noise shaping network (FiNS) \cite{FiNS}, wherein the late  reverberation content is modeled via a combination of noise bases; however, we propose an alternate estimation method that explicitly leverages the goals of a speech recognizer. Additionally, while room acoustics-agnostic domain adaptation strategies have been explored to handle the domain mismatch issue \cite{domain_adapt}, we draw and study an explicit connection between improved RIR estimation and improved ASR performance.

Overall, we propose an approach for speech-to-RIR estimation that directly benefits ASR. Our main contributions are twofold: 
We first propose a novel GAN-based RIR estimator (S2IR-GAN) using an energy-decay-relief loss, which we expect to capture the energy-based acoustic properties of input reverberant speech, and a discriminator loss to estimate the fine structure of the RIR. We demonstrate $22\%$ improvement in capturing early reflection energy (ERE) and $17\%$ improvement in energy decay relief (EDR). 
We then evaluate the benefits of the proposed model on an ASR task and demonstrate a $6.9\%$ reduction of word error rate relative to existing RIR estimators when handling reverberant speech. 

\section{Related Work}
\label{sec:related}


Several algorithms have been proposed to blindly estimate an RIR from a reverberant source signal using traditional signal-processing approaches~\cite{crammer2006room,lin2006bayesian,es1,es4,es2}. For some methods that take an $\ell_{1}$-norm-based approach, performance depends significantly on the choice of a regularization parameter corresponding to a real-world scenario \cite{es2}; some require multi-channel speech signals ~\cite{es2}; and most assume that either the source signal is a modulated Gaussian pulse~\cite{es1,es4}, or that the speaker and microphone characteristics are known~\cite{crammer2006room,lin2006bayesian}. For far-field ASR tasks, however, we are required to estimate RIRs from reverberant speech source signals independent of speaker and microphone characteristics. 


Recently, a neural network model was proposed to estimate the RIR from single-channel reverberant speech (FiNS) \cite{FiNS}. The FiNS model directly estimates early RIR components, and estimates late components as a combination of decaying filtered noise signals. In contrast, we propose a GAN-based architecture (S2IR-GAN) to directly estimate the entire RIR up to 0.25 seconds. RIRs with full duration and RIRs cropped to a duration of 0.25 seconds are known to give similar performance in far-field ASR tasks in meeting rooms~\cite{fast-rir}. 

We explicitly train our S2IR-GAN on an energy-based cost function to estimate an RIR with an energy distribution similar to the ground-truth RIR. Estimating energy-based acoustic parameters such as reverberation time ($T_{60}$) and direct-to-reverberant ratio ($DRR$) and incorporating them in speech dereverberation and speech recognition systems have shown improved performance~\cite{reverbaware1,reverbaware2,reverbaware3}. Therefore, we expect that accurately estimating the energy distribution in the estimated RIRs helps the S2IR-GAN model improve its performance over other approaches in far-field ASR tasks.

\vspace{-0.2cm}

\section{Our Approach}
\label{sec:approach}

\subsection{RIR Estimation from Reverberant Speech}

We alternatingly train our RIR estimation network ($E_{Net}$) and a discriminator network ($D_{Net}$) using reverberant speech ($S_{R}$) and the corresponding RIR ($R_{G}$) in the data distribution $p_{d}$ at each iteration. Our RIR estimation network estimates the RIR of the input reverberant speech and our discriminator network is optimized to differentiate the estimated RIR from the ground truth RIR. The objective function of $E_{Net}$ consists of the EDR error (see below), the modified conditional GAN (CGAN)~\cite{fast-rir} error, and the mean square error (MSE). We use a modified CGAN objective function to train $D_{Net}$. In the modified CGAN, both $E_{Net}$ and $D_{Net}$ are conditioned on the reverberant speech ($S_{R}$) to estimate a single precise RIR for the given $S_{R}$.

\textbf{EDR Error:}
The EDR describes the energy remaining in the RIR in a specific frequency band centered at $b_k$ Hz at time $t$ seconds. In the following EDR equation, $H(r,t,k)$ is the bin $k$ of the short-time Fourier transform of the RIR $r$ at time $t$. The total number of time frames is $T$.
{
\belowdisplayskip 0.5\belowdisplayshortskip
\small
\begin{equation}\label{EDRelief}
\begin{aligned}[b]
  EDR(r,t,b_k)  = \sum_{t=t}^T |H(r,t,k)|^2.
\end{aligned}
\end{equation}
} 


\noindent We calculate the $EDR$ of the estimated RIR using $E_{Net}$ and the ground truth RIR ($R_{G}$) at a set of octave frequency bands (B) with center frequencies from 16Hz to 4000 Hz. In Eq.~\ref{edr_loss}, $\mathbb{E}$ is the expectation. We calculate the EDR loss as follows:
{
\belowdisplayskip 0.4\belowdisplayshortskip
\small
\begin{equation}\label{edr_loss}
\begin{aligned}[b]
    \mathcal{L}_{EDR} = \mathbb{E}_{(S_{R}, R_{G}) \sim p_{d}} [\mathbb{E}[(EDR(E_{Net}(S_{R}),t,b_{k})
    \\ - EDR(R_{G},t,b_{k}))^{2}]].
\end{aligned}
\end{equation}
}


\noindent Our EDR loss helps the RIR estimator to capture energy-based acoustic properties of the  RIR~\cite{sagnik}. 

\textbf{CGAN Error (RIR Estimator):} 
The CGAN error is used to estimate the RIR from the reverberant speech ($S_{R}$) using our RIR estimator ($E_{Net}$) that is difficult to differentiate from the ground truth RIR by the $D_{Net}$ during training. 

{
\belowdisplayskip 0.4\belowdisplayshortskip
\small
\begin{equation}\label{CGAN_loss}
\begin{aligned}[b]
    \mathcal{L}_{CGAN} = \mathbb{E}_{S_{R} \sim p_{d}}[\log(1 - D_{Net}(E_{Net}(S_{R}),S_{R}))].
\end{aligned}
\end{equation}
\vspace{-0.3cm}
} 

\textbf{MSE:}
For each reverberant speech example ($S_{R}$), we calculate the squared difference of each time sample ($t$) in the estimated RIR $E_{Net}(S_{R})$ and the ground truth RIR $R_{G}$.
{\belowdisplayskip 0.6\belowdisplayshortskip
\small
\begin{equation}\label{mse_loss}
\begin{aligned}[b]
    \mathcal{L}_{MSE} = \mathbb{E}_{(S_{R}, R_{G}) \sim p_{d}}[\mathbb{E}[(R_{G}(S_{R},t) - E_{Net}(S_{R},t))^{2}]].
\end{aligned}
\end{equation}
\vspace{-0.3cm}
}

\noindent We alternatingly train $E_{Net}$ to minimize the objective function $\mathcal{L}_{E_{Net}}$ (Eq.~\ref{generator_loss}) and $D_{Net}$ (Eq.~\ref{discriminator_loss}) to maximize the objective function $\mathcal{L}_{D_{Net}}$. $\mathcal{L}_{D_{Net}}$ is maximized to differentiate $R_{G}$ from $E_{Net}(S_{R})$ during training by $D_{Net}$. We use the weights $\lambda_{EDR}$ and $\lambda_{MSE}$ to control the contribution of $\mathcal{L}_{EDR}$ and $\mathcal{L}_{MSE}$ respectively in $\mathcal{L}_{E_{Net}}$:

{
\belowdisplayskip 0.4\belowdisplayshortskip
\small
\begin{equation}\label{generator_loss}
\begin{aligned}[b]
    \mathcal{L}_{E_{Net}} = \mathcal{L}_{CGAN} + \lambda_{EDR} \; \mathcal{L}_{EDR} + \lambda_{MSE} \; \mathcal{L}_{MSE},
\end{aligned}
\end{equation}
\vspace{-0.5cm}
}

{
\belowdisplayskip 0.4\belowdisplayshortskip
\small
\begin{equation}\label{discriminator_loss}
\begin{aligned}[b]
    \mathcal{L}_{D_{Net}} = \mathbb{E}_{(R_{G},S_{R}) \sim p_{d}}[\log(D_{Net}(R_{G}(S_{R}),S_{R}))] \\
    + \mathbb{E}_{S_{R} \sim p_{d}}[\log(1 - D_{Net}(E_{Net}(S_{R}),S_{R}))].
\end{aligned}
\end{equation}
\vspace{-0.5cm}
}

\begin{figure}[t] 
	\centering
	\includegraphics[width=0.85\linewidth]{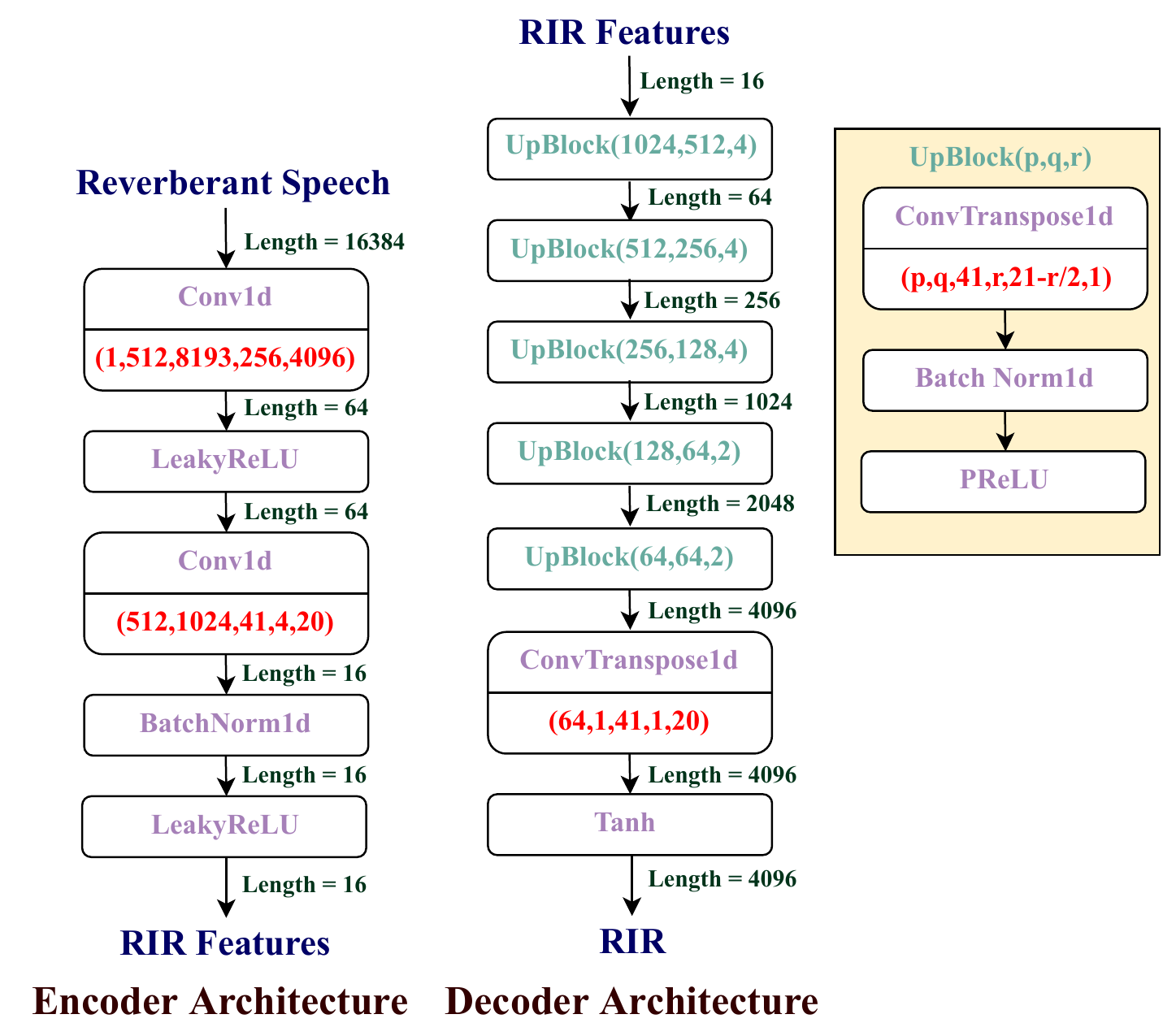}
	\caption{The encoder-decoder architecture of our S2IR-GAN. Our encoder network extracts the RIR features, and our decoder network constructs an RIR from the extracted features. For Conv1d layers and ConvTranspose1d layers, the parameters are input channels, output channels, kernel size, stride and padding. The last parameter of ConvTranspose1d is output padding. We use a negative slope of 0.2 for the Leaky ReLU layers.} 
	\label{architecture}
	\vspace{-0.5cm}
\end{figure}

\subsection{Network Architecture}

\textbf{RIR Estimator:}
We propose an encoder-decoder architecture to estimate the RIR from the reverberant speech with a sampling rate of 16000 Hz (Fig.~\ref{architecture}). The reverberant speech ($S_R$) can be described as a convolution of clean speech ($S_C$) and an RIR as follows:
{\small
\belowdisplayskip 0.6\belowdisplayshortskip
\begin{equation}\label{convolution}
\begin{aligned}[b]
   S_R[n] = \sum_{s=1}^S RIR[s] * S_C[n-s],
\end{aligned}
\end{equation}
}
\noindent where $S$ is the total number of samples in the RIR. To capture the features of the RIR of length 4096 from the reverberant speech, we use a large convolutional layer of length 8193 in the first layer. Later, we reduce the dimension of the extracted RIR features from 64 to 16 while increasing the number of channels from 512 to 1024.

We construct the RIR from the encoded features in the decoder network. We use 5 sets of transpose convolution layers, batch normalization and PReLU layers to gradually increase the length from 16 to 4096 and reduce the number of channels from 1024 to 64. We use a transpose convolution layer with stride 1 to collapse the number of channels from 64 to 1 and obtain the weighted averaged estimated RIR. In the last layer, we use the tanh activation function because the RIR contains both negative and positive values. Fig.~\ref{architecture} shows our encoder-decoder architecture in detail. We adapt the discriminator ($D_{Net}$) network architecture from FAST-RIR~\cite{fast-rir}. We pass the 1st 512 samples of the $S_R$ as the condition to $D_{Net}$.

\subsection{Dataset and Training}
\label{subsec:training_data}
We generate 98,316 one-second duration synthetic training examples of reverberant speech by convolving a subset of 360 hours of clean speech (train-clean-360) from the LibriSpeech dataset~\cite{LibriSpeech} with RIRs from the Meta RIR (MRIR) dataset. To pick the best model during training, we use the validation EDR loss. We trained our network for 200 epochs using the RMSprop optimizer. We used a batch size of 128 and an initial learning rate of 8 x $10^{-5}$. The learning rate is decayed by 0.7 after every 40 epochs. 
\vspace{-0.2cm}



\begin{table*}[t]
	\renewcommand{\arraystretch}{0.70} 
	\caption{The log of Energy decay relief (EDR) loss (Eq.~\ref{edr_loss}) and the mean absolute error of the early reflection energy (ERE loss). In this table, we compare the baseline model, FiNS~\cite{FiNS} and our S2IR-GAN. The best results are shown in \textbf{bold}.}
	\label{table_loss}
	\centering
	\begin{tabular}{@{}llcccccccccc@{}}	
		\toprule
		 \textbf{Loss} & \textbf{Method}
			& \multicolumn{9}{c}{\textbf{Frequency}}\\
\cmidrule(r{1pt}){3-11} 
		&	&\textbf{16Hz} &\textbf{32Hz}	&\textbf{63Hz} &\textbf{125Hz}	&\textbf{250Hz} &\textbf{500Hz}	&\textbf{1000Hz} &\textbf{2000Hz}&\textbf{4000Hz} &\\  
		\midrule
	\textbf{Log}	&Baseline & -0.08 & -4.77& -5.36 & -4.11& -2.66 & -2.85& -1.08 & -2.80& -1.00 \\
	\textbf{EDR}	&FiNS & -9.10 & -9.80& -9.34&-4.17& -3.00 & -3.64& -3.16 & -3.00& -1.48 \\
		\textbf{Loss}&\textbf{Ours}&  \textbf{-10.74} & \textbf{-10.72} & \textbf{-9.47} & \textbf{-4.70}& \textbf{-3.42} & \textbf{-4.23}& \textbf{-3.85} & \textbf{-3.66}& \textbf{-2.08} \\
		\midrule
	\textbf{ERE}	&Baseline &    8.75 & 7.66 & 4.47 & 13.31 & 11.41 & 6.44 & 5.65 & 7.67 & 5.80 \\
	\textbf{Loss}	&FiNS&    7.41 & 6.89 & 4.69 & 3.77 & 4.21 & 3.62 & 3.81 & 3.75 & 3.67  \\
		&\textbf{Ours} &    \textbf{5.88} & \textbf{5.33} & \textbf{3.34} & \textbf{3.27} & \textbf{3.11} & \textbf{2.88} & \textbf{2.87} & \textbf{2.83} & \textbf{2.91} \\

		\bottomrule
	\end{tabular}
	\vspace{-0.4cm}
\end{table*}

\begin{table}[t]
	\renewcommand{\arraystretch}{0.70} 
	\caption{MSE (Eq.~\ref{mse_loss}) and DRR error of the estimated RIRs from the baseline model, FiNS~\cite{FiNS} and S2IR-GAN.}
	\label{table_mse}
	\centering
	\begin{tabular}{@{}lllc@{}}	
		\toprule
		\multicolumn{2}{l}{{\textbf{METHOD}}}
			& \textbf{MSE} &\textbf{DRR (dB)}\\
		
		\midrule
		&Baseline    & \textbf{3.0x\boldmath$10^{-4}$} & 6.63\\
		&FiNS  & 3.7x$10^{-4}$  & 3.31\\
		&\textbf{S2IR-GAN (ours)} &\textbf{3.0x\boldmath$10^{-4}$} & \textbf{3.28}\\
			
		\bottomrule
	\end{tabular}
	\vspace{-0.3cm}
\end{table}

\section{Acoustic Evaluation}
\label{sec:evalaute}

We evaluate the performance of our proposed approach using EDR loss (Eq.~\ref{edr_loss}), DRR error, early reflection energy (ERE) loss and MSE (Eq.~\ref{mse_loss}) of the estimated RIR and the ground truth RIR. Finally, we also compare the ground truth RIR and the estimated RIR using our S2IR-GAN in the time domain.

The EDR of the RIR contains enough information to construct an equivalent RIR with the same gross temporal and spectral features while the fine structure can be different~\cite{EDC}. Acoustic parameters such as $T_{60}$ and early decay time are measured on the slope of EDR. Therefore, we use EDR loss for evaluation.  DRR, the log ratio of the energy of the direct sound and the sound arriving after it, is an acoustic parameter often used in ASR applications to measure the amount of distortion introduced into speech by the RIR~\cite{drruse}; DRR estimation is also directly integrated into some ASR systems~\cite{reverbaware2}.  Here, we compute the DRR error as the mean absolute error (MAE) of the DRR between the estimated and the ground-truth RIRs.  We also include the ERE loss, which is the MAE of the early reflection energy between the estimated and the ground truth RIR. ERE measures the total early sound energy between 0 and 80ms:


{\small
\belowdisplayskip 0.5\belowdisplayshortskip
\begin{equation}\label{clarity_loss}
\begin{aligned}[b]
   ERE =10\log{ \sum_{t=0}^{80 ms} |RIR(t)|^2}.
\end{aligned}
\end{equation}
\vspace{-0.3cm}
}

We created 5000 reverberant test examples by convolving clean speech from the LibriSpeech dataset (train-clean-100) with RIRs from the MRIR dataset not used for training. We compared our performance against a baseline model and the state-of-the-art RIR estimator (FiNS). We modified the input and output dimensions of the FiNS model to match our S2IR-GAN. For a fair comparison, we have trained FiNS and our S2IR-GAN using the same training data (Section~\ref{subsec:training_data}). We use a pre-trained open-source speech enhancement toolkit named ESPNET-SE~\cite{espnet} as a baseline model.


\textbf{Baseline:} ESPNET-SE can perform speech dereverberation and denoising. We pass the reverberant speech test data as the input to the network and get clean speech as the output. We compute the RIR using the reverberant speech ($S_R$) and the corresponding output clean speech ($S_C$) as follows:
{
\belowdisplayskip 0.4\belowdisplayshortskip
\small
\begin{equation}\label{espnet}
\begin{aligned}[b]
  RIR = \mathcal{F}^{-1}\left(\frac{\mathcal{F}(S_R)}{\mathcal{F}(S_C)}\right),
\end{aligned}
\end{equation}
\vspace{-0.3cm}
}

\noindent where $\mathcal{F}$ and $\mathcal{F}^{-1}$ are the Fourier transform and inverse Fourier transform, respectively.
\begin{figure}[!h] 
\centering
\subfloat[250 Hz.]{\includegraphics[width=0.47\columnwidth]{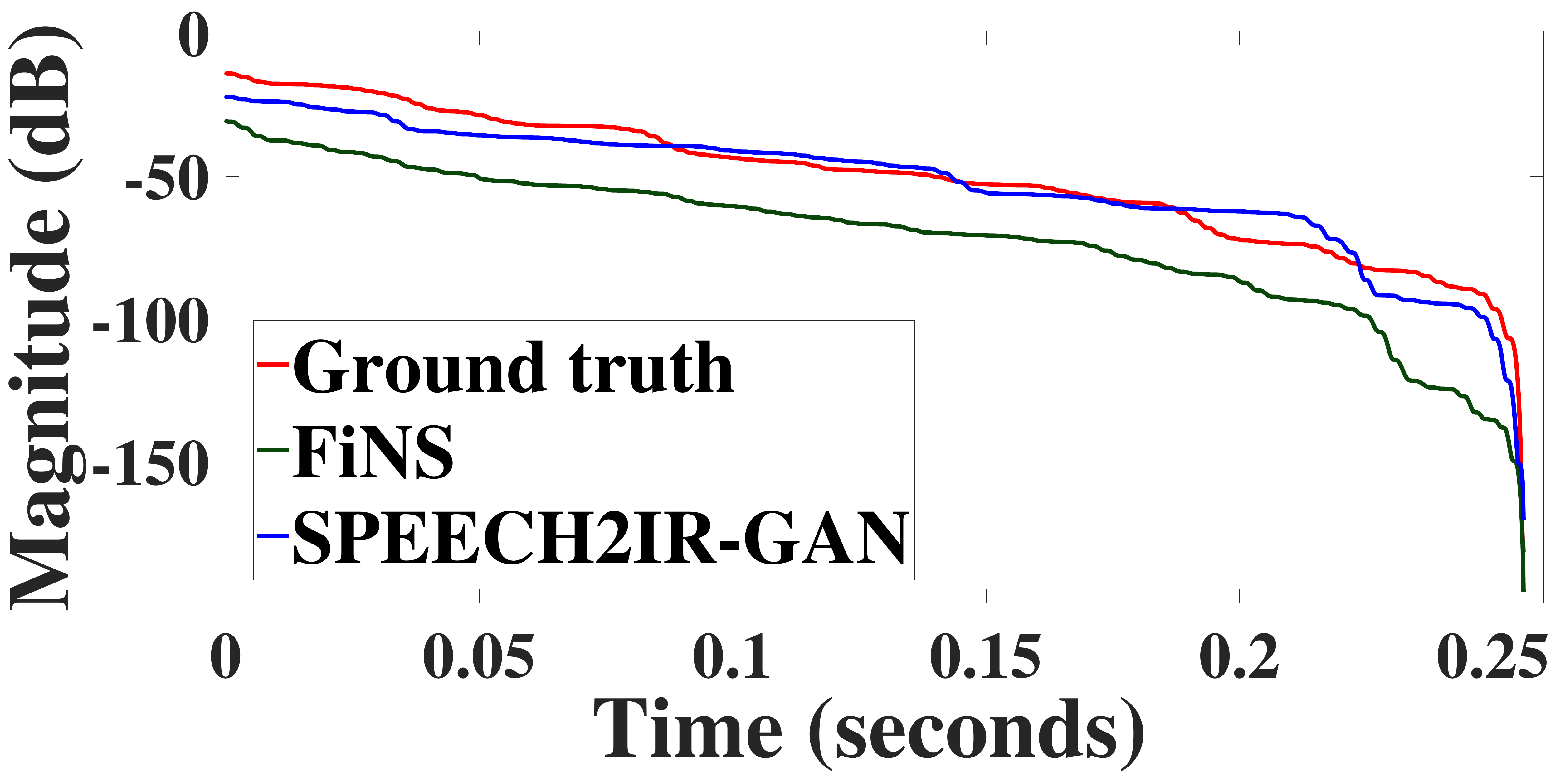}}
\quad
\subfloat[2000 Hz.]{\includegraphics[width=0.47\columnwidth]{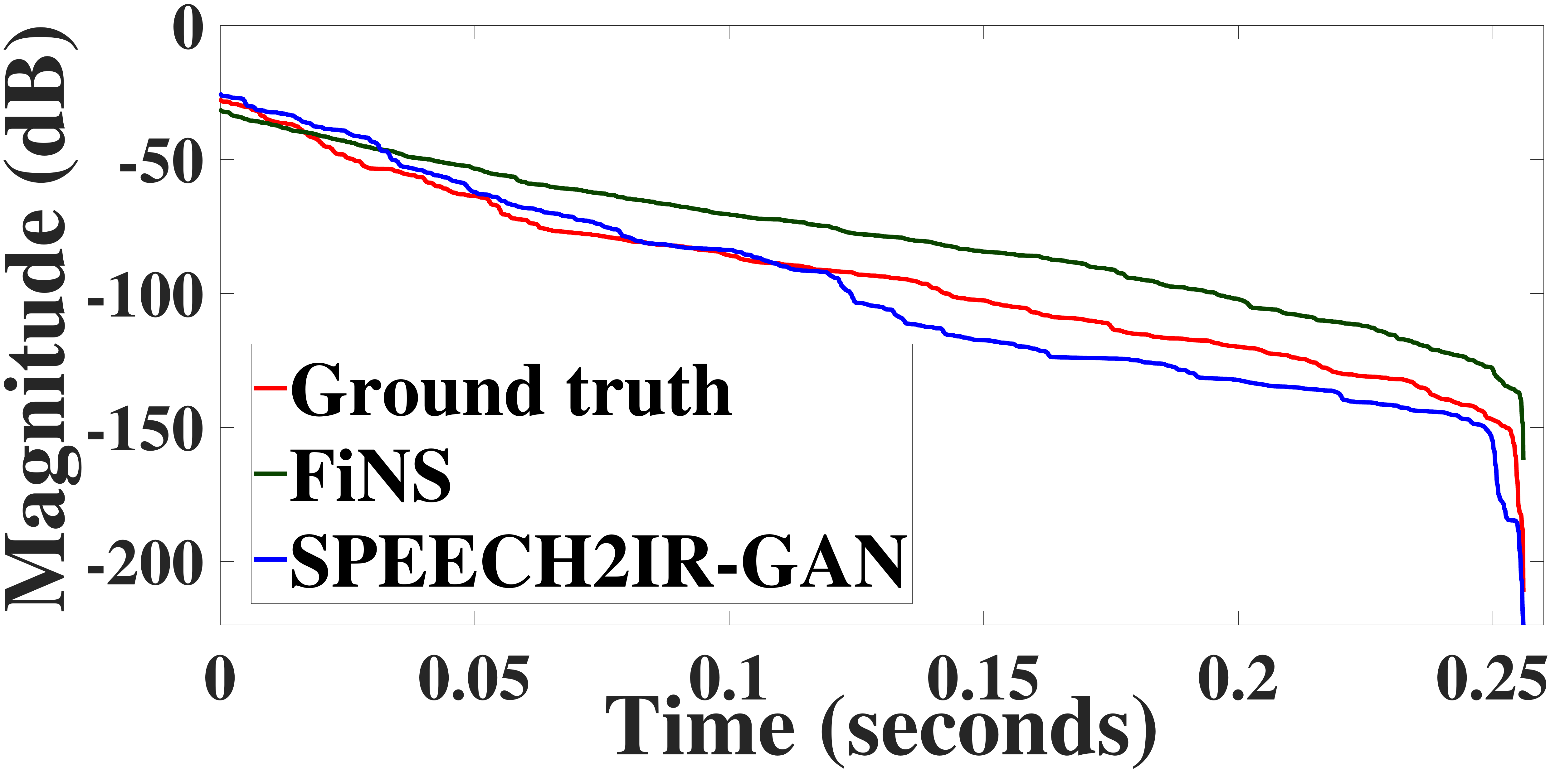}}
\caption{The EDR (Eq.~\ref{EDRelief}) of the ground truth RIR, and the estimated RIR using FiNS and our S2IR-GAN models at 250 HZ and 2000 Hz. We can see that the EDR of S2IR-GAN is closest to the ground truth EDR. }
\label{edr_curve}
\vspace{-0.3cm}
\end{figure}

\begin{figure}[h]
  \centering
  \includegraphics[width=0.8\linewidth]{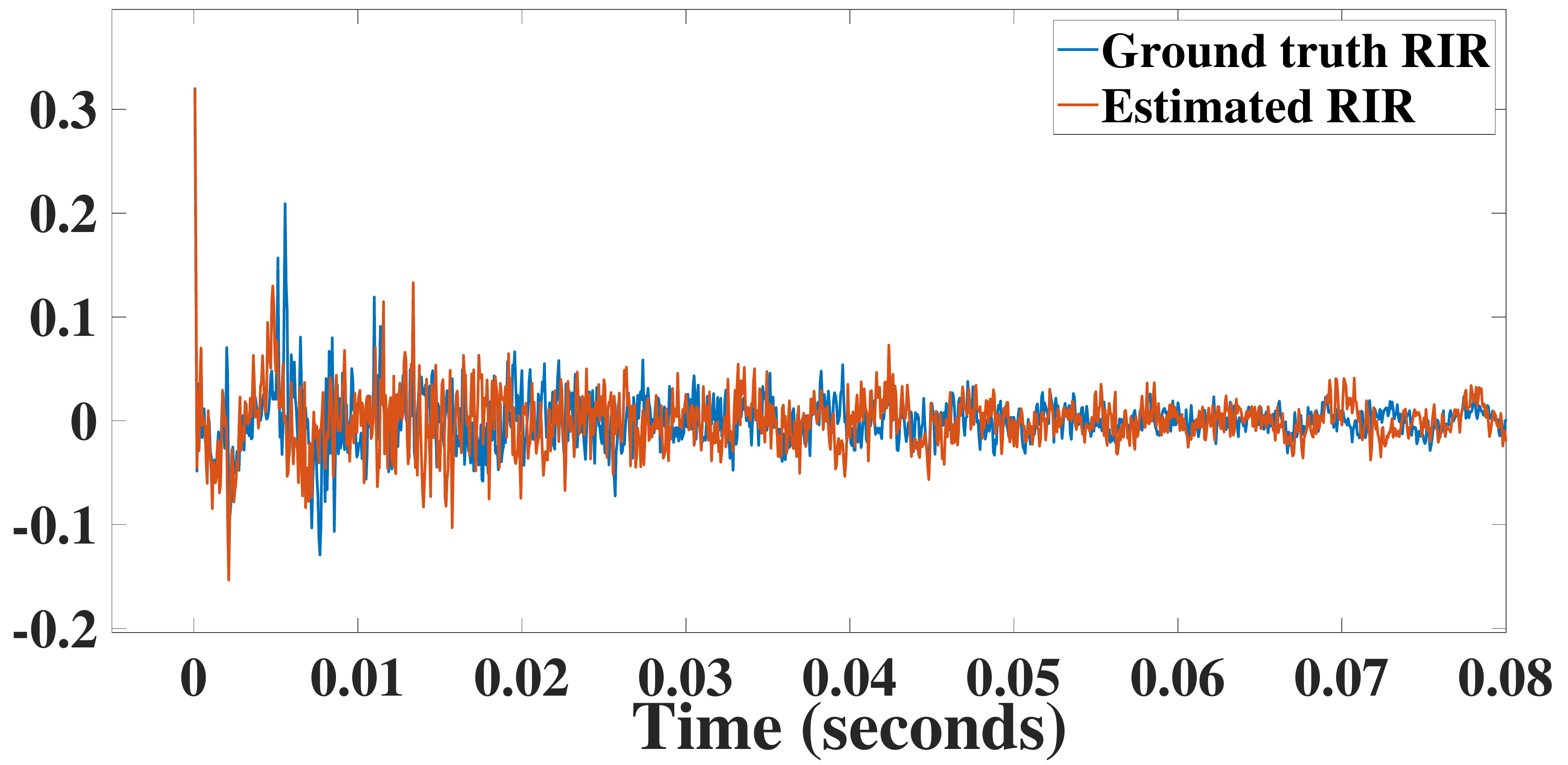}
  \caption{Time domain plot of the ground truth RIR and the estimated RIR from our S2IR-GAN. The estimated RIR has similar macroscopic structure as the ground truth RIR while the fine structure differs from the ground truth RIR. } 
   
  \label{rir_plot}
\vspace{-0.3cm}
\end{figure}

Table~\ref{table_loss} shows the log of EDR loss and ERE loss for the baseline model, FiNS and our S2IR-GAN. We calculate the loss over the octave-bands with center frequencies up to 4000 Hz. We can see that on average, our proposed S2IR-GAN outperforms the FiNS by 17\% in the log of EDR loss and 22\% in ERE loss. In Fig.~\ref{edr_curve}, we can see the ground truth EDR and the EDR of the estimated RIRs using FiNS and S2IR-GAN at 250 Hz and 2000 Hz. The EDR of our S2IR-GAN is closest to the ground truth EDR. 

We calculated the MSE and DRR error of the estimated RIRs for 3 different methods (Table~\ref{table_mse}). We can see that Baseline and S2IR-GAN give the lowest MSE error. Baseline gives higher DRR error when compared to S2IR-GAN. Fig.~\ref{rir_plot} shows an example of the ground truth RIR and the estimated RIR from our S2IR-GAN. We can see that our network can estimate general patterns in the ground truth RIR. However, our network cannot accurately estimate the fine temporal  structure of the ground truth RIR. The estimated RIRs are perceptually similar to the ground truth RIR, as can be heard through audio samples of ground truth RIRs and estimated RIRs\footnote{\url{https://anton-jeran.github.io/S2IR/}}.
\vspace{-0.3cm}

\section{ASR Evaluation}
\label{sec:results}
We evaluate the performance of our S2IR-GAN in the Kaldi automatic speech recognition (ASR) experiment\footnote{\url{https://github.com/RoyJames/kaldi-reverb/}}. We use close-talk speech data (IHM) and far-field speech data (SDM) in the AMI corpus~\cite{ami} for our experiment. From the SDM corpus, we sample 2000 reverberant speech examples of approximately one second duration. We input the sampled reverberant speech to the FiNS model and our proposed S2IR-GAN and estimate the RIRs from the input speech. The close-talk IHM speech is convolved with the estimated RIRs to create synthetic reverberant speech training data.

We train the modified Kaldi ASR recipe with the synthetic training data and test the ASR model on the real-world reverberant SDM data. We also train the ASR model with unmodified IHM data as our baseline model. We use word error rate (WER) to evaluate the performance of the ASR system. Lower WER indicates that the reverberation effects in the training speech data are closer to the test speech data (SDM)~\cite{gwa}. From Table~\ref{table_ami}, we can see that our S2IR-GAN outperforms FiNS by 6.9\%.

\vspace{-0.3cm}

\begin{table}[t]
  \setlength{\tabcolsep}{1.8pt}
	\renewcommand{\arraystretch}{0.7} 
	\caption{Far-field ASR results were obtained for far-field speech data recorded by single distance microphones (SDM) in the AMI corpus. The best results are shown in \textbf{bold}.}
	\label{table_ami}
	\centering
	\begin{tabular}{@{}llc@{}}	
		\toprule
		\multicolumn{2}{l}{{\textbf{Training Dataset}}}
			& \textbf{Word Error Rate} [\%]\\
		&\textbf{Clean Speech $\circledast$ RIR} &\\  
		\midrule
		&IHM $\circledast$ None  & 64.2 \\
		&IHM $\circledast$ FiNS  & 60.9 \\
		&\textbf{IHM \boldmath$\circledast$ S2IR-GAN (ours)} & \textbf{54.0} \\
			
		\bottomrule
	\end{tabular}
	\vspace{-0.3cm}
\end{table}

\section{Discussion}

\label{sec:discuss}
In this work, we present a method for the improved estimation of RIRs in the context of far-field speech recognition.  Our S2IR-GAN model outperforms the state-of-the-art RIR estimator (FiNS) on acoustic metrics such as energy decay relief loss, early reflection energy loss, DRR error, and MSE. It is also shown to outperform FiNS in a downstream ASR task.
The main limitation of this work is that the network cannot capture the fine temporal structure of an RIR from reverberant speech, though the perceptual implications of this shortcoming are unclear.  In the future, we would like to extend this work to improve performance of the current model by expanding the set of RIRs used to augment training data; consider RIR estimation while leveraging the goals of other downstream tasks such as dereverberation or source separation; or estimate RIRs from a larger set of input modalities, such as multi-channel speech signals or audio-visual data.




\bibliographystyle{IEEEbib}
\bibliography{strings,refs}

\end{document}